\documentclass[final, 10pt]{IEEEtran}
\IEEEoverridecommandlockouts
\usepackage{cite}
\usepackage{amsmath,amssymb,amsfonts}
\usepackage{algorithmic}
\usepackage{graphicx}
\usepackage{textcomp}
\usepackage{xcolor}
\usepackage{enumitem}
\usepackage{float}
\def\BibTeX{{\rm B\kern-.05em{\sc i\kern-.025em b}\kern-.08em
    T\kern-.1667em\lower.7ex\hbox{E}\kern-.125emX}}

\floatstyle{plain}
\newfloat{copyrightbox}{!hb}{loc}
\floatname{copyrightbox}{Copyright Notice}

\newcommand\copyrighttext{%
\footnotesize 
\textit{This is a preprint accepted at the 10th IEEE International Conference On Data Science And Advanced Analytics (DSAA 2023). The conference proceedings will be published by the IEEE Xplore Digital Library with ISBN: 979-8-3503-4503-2.\textcopyright IEEE.}
}

\begin{document}

\title{Exploring Deep Learning for Full-disk Solar Flare Prediction with Empirical Insights from Guided Grad-CAM Explanations\\
}
\author{\IEEEauthorblockN{Chetraj Pandey\textsuperscript{1}, Anli Ji\textsuperscript{1}, Trisha Nandakumar\textsuperscript{2}, Rafal A. Angryk\textsuperscript{1}, Berkay Aydin\textsuperscript{1}}\\
\IEEEauthorblockA{\textit{\textsuperscript{1,2}Dept. of Computer Science, Georgia State University, Atlanta, GA, USA} \\
\textit {\textsuperscript{1}\{cpandey1, aji1, rangryk, baydin2\}@gsu.edu}}\\
\textit{\textsuperscript{2}trishan104@gmail.com}}

\maketitle
\begin{abstract}
This study progresses solar flare prediction research by presenting a full-disk deep-learning model to forecast $\geq$M-class solar flares and evaluating its efficacy on both central (within $\pm$70$^\circ$) and near-limb (beyond $\pm$70$^\circ$) events, showcasing qualitative assessment of post hoc explanations for the model's predictions, and providing empirical findings from human-centered quantitative assessments of these explanations. Our model is trained using hourly full-disk line-of-sight magnetogram images to predict $\geq$M-class solar flares within the subsequent 24-hour prediction window. Additionally, we apply the Guided Gradient-weighted Class Activation Mapping (Guided Grad-CAM) attribution method to interpret our model's predictions and evaluate the explanations. Our analysis unveils that full-disk solar flare predictions correspond with active region characteristics. The following points represent the most important findings of our study: (1) Our deep learning models achieved an average true skill statistic (TSS) of $\sim$0.51 and a Heidke skill score (HSS) of $\sim$0.38, exhibiting skill to predict solar flares where for central locations the average recall is $\sim$0.75 (recall values for X- and M-class are 0.95 and 0.73 respectively) and for the near-limb flares the average recall is $\sim$0.52 (recall values for X- and M-class are 0.74 and 0.50 respectively); (2) qualitative examination of the model's explanations reveals that it discerns and leverages features linked to active regions in both central and near-limb locations within full-disk magnetograms to produce respective predictions. In essence, our models grasp the shape and texture-based properties of flaring active regions, even in proximity to limb areas—a novel and essential capability with considerable significance for operational forecasting systems.
\end{abstract}

\begin{copyrightbox}
\centering
  \fbox{\parbox{\dimexpr\linewidth-5\fboxsep-3\fboxrule\relax}{\copyrighttext}}
\end{copyrightbox}
\section{Introduction}
Solar flares are transient but intense outbursts of energy that emanate from the Sun's surface in the form of extreme ultraviolet and X-ray radiation and are one of the central phenomena in space weather forecasting. Flares are typically classified in five different classes based on their peak X-ray flux by National Oceanic and Atmospheric Administration (NOAA) \cite{spaceweather}, where flare intensities are characterized by a major class (i.e., A, B, C, M, and X) and a numerical strength value within a class (from 1.0 to 9.9). These indicate the order of magnitude of the peak X-ray flux in a logarithmic scale (e.g., X-ray flux values for B3.7 is $3.7\times 10^{-7}Wm^{-2}$, C7.2 is $7.2\times 10^{-6}Wm^{-2}$, M1.4 is $1.4\times 10^{-5}Wm^{-2}$, or X2.1 is $2.1\times 10^{-4}Wm^{-2}$).
M- and X-class solar flares are relatively scarce events and significantly more powerful than the other flare classes which garner the attention of researchers due to their potential to cause near-Earth impacts that pose a substantial risk to both space-based and terrestrial infrastructures, disrupting satellite communications, power grids, and aviation, making solar flare prediction a complex and critical area of research in space weather forecasting \cite{Yasyukevich2018}. 

Active regions (ARs) on the Sun represent areas where the sun's magnetic field experiences disruptions, leading to a variety of solar phenomena such as solar flares, coronal mass ejections (CMEs), and solar energetic particle (SEP) events. The majority of solar flare prediction models concentrate on these regions of interest, offering forecasts for each AR, as they are primarily responsible for space weather events. To generate a comprehensive forecast using an AR-centric model, the likelihood of a flare occurring within each AR ($P_{FL}(AR_i)$) is usually consolidated utilizing a heuristic function as described in \cite{Pandey2022f}. The equation for this aggregation is $P_{aggregated} = 1 - \prod_{i}\big[1-{P_{FL}(AR_i)\big]}$, which computes the probability of having a minimum of one flaring AR, under the assumption that flares from different ARs occur independently and infers that all ARs hold equal significance in the combination. This constrains an accurate assessment of the full-disk flare prediction probability. 

Furthermore, the magnetic field measurements, which underpin AR-based forecasting techniques, are influenced by projection effects when ARs are in proximity to the limbs\cite{Falconer2016}. Consequently, the data is restricted to ARs situated within a range of $\pm$30$^{\circ}$ \cite{Huang2018} to $\pm$70$^{\circ}$ \cite{Ji2020} of the solar disk due to significant projection effects \cite{Hoeksema2014}. The full-disk models, however, utilize magnetograms corresponding to the entire solar disk to rely on shape-based features such as size, directionality, shapes of sunspots, and polarity inversion lines similar to the findings from \cite{Korss2014, McCloskey2016, Deshmukh2020, Ji2023}. While projection effects persist in these images, we show that the full-disk models are capable of learning and predicting flare productivity of areas close to the limb of the Sun, offering an essential component for operational systems. 

Recently, many studies (e.g., \cite{Nishizuka_2017, Nishizuka2018, Huang2018, Li2020, Pandey2021, Pandey2022, Pandey2022f, Whitman2022, Hong2023}) have led to the successful application of machine learning and deep learning methods in predicting solar flares. These methods have shown promising experimental results, indicating their potential to improve forecasting accuracy in the field of solar flare prediction. However, deep learning models are often considered black-box models due to their complex data representations, making it challenging to understand the reasoning behind their predictions. This lack of transparency can pose problems in critical applications like solar flare prediction, where the model's reliability is crucial. The absence of transparency in the decision-making process lowers confidence in the accuracy and reliability of the predictions made. Post hoc explanations offer insights into a model's decision-making process, enabling the identification of potential errors and biases in the data, and subsequent improvements to the model's performance and reliability, which addresses the issue of transparency and makes it more trustworthy for critical applications. While post hoc explanations effectively highlight crucial input features and enhance trust, a gap persists in our understanding due to the inherent opacity of these models' decision-making processes. Therefore, an approach for meaningful translations remains essential for fully understanding these explanations. This paper serves as an initial step in shedding light on our models' effectiveness by clarifying the model's reasoning with post hoc explanations and enhancing confidence in the model's reliability and accuracy, especially in cases where incorrect predictions could result in substantial consequences, such as in space weather forecasting.

The focus of this paper is to study whether full-disk models for solar flare prediction can be relied upon for operational forecasting applications as this exploration extends to the prediction of near-limb flare events with model explanations, which hold a notably critical role. This paper builds upon the convolutional neural network (CNN) based model in \cite{pandeyds2023} to predict $\geq$M-class flares, trained with compressed full-disk line-of-sight magnetograms. The novel contributions of this paper are as follows: (i) We highlight the overall improved performance of our solar flare prediction model exhaustively with quantitative analysis of predictive performance on near-limb and central locations, (ii) We utilized a recent attribution method to explain and interpret the decisions of our deep learning model and present a case study that shows the explanations of model's decision in spatiotemporal progression, and (iii) More importantly, for the first time, we provide empirical evidence that our models can tackle the prediction of flares appearing on near-limb regions of the Sun by analyzing the explanations with quantitative human-centered approach \cite{explainai}.

The remainder of the paper is structured as follows: Section~\ref{sec:rel} provides an overview of existing studies on machine learning and deep learning-based data-driven solar flare prediction models. In Section~\ref{sec:data}, we detail the process of data collection with labeling and explain the subsequent data distribution. In Section~\ref{sec:modelexp}, we outline our methodology by describing the architecture of our flare prediction model and providing a detailed explanation of the method used for explanation. In Section~\ref{sec:expt}, we present our experimental design and findings from the model evaluation. Furthermore, we offer case-based qualitative interpretations of explanations for limited instances and a quantitative analysis of explanations for entire X-class flares in our dataset. Finally, in Section~\ref{sec:conc}, we summarize our findings and suggest avenues for future research.

\section{Related Work}\label{sec:rel}
Currently, to the best of our knowledge, four strategies are employed for flare prediction: (i) empirical human prediction (e.g., \cite{Crown2012, Devos2014}), (ii) statistical prediction (e.g., \cite{Lee2012, Leka2018}), (iii) physics-based numerical simulations (e.g., \cite{Kusano2020, Korss2020}), and (iv) machine learning and deep learning approaches (e.g., \cite{Bobra2015, Huang2018, Li2020, Pandey2021, Pandey2022, Pandey2022f, Whitman2022, Hong2023}). With the rapid progress in machine learning and deep learning techniques, their application to solar flare prediction has significantly expedited research efforts. For instance, in \cite{Nishizuka2018}, a multi-layer perceptron-based model was employed for predicting $\geq$C- and $\geq$M-class flares. The model utilized 79 manually selected physical precursors extracted from multi-modal solar observations.

Deep learning models have recently emerged as a prominent choice for solar flare prediction. In \cite{Huang2018}, a CNN-based model was trained using AR patches extracted from line-of-sight magnetograms within $\pm$30$^{\circ}$ of the central meridian to predict $\geq$C-, $\geq$M-, and $\geq$X-class flares. Similarly, \cite{Li2020} developed a CNN-based model that issued binary class predictions for $\geq$C- and $\geq$M-class flares within 24 hours using Space-Weather Helioseismic and Magnetic Imager Active Region Patches (SHARP) data \cite{Bobra2014}. The SHARP data was extracted from solar magnetograms using AR patches located within $\pm45^{\circ}$ of the central meridian. Notably, both models had limited operational capability, as they were restricted to small portions of the observable disk in central locations ($\pm30^{\circ}$ and $\pm45^{\circ}$). 

In the context of explainability in solar flare prediction models,  \cite{Bhattacharjee2020} used an occlusion-based method to interpret a CNN-based solar flare prediction model trained with AR patches. Similarly, \cite{Yi2021} presented visual explanation methods for a deep learning-based flare prediction model. They used Grad-CAM \cite{gradcam}, and Guided Backpropagation \cite{gbackprop} to show the relationship of physical parameters of ARs with flare activity in the context of the daily occurrence of C-, M-, and X-class flares. They used daily observations of solar full-disk line-of-sight magnetograms at 00:00 UT, and their models show limitations for the near-limb flares. Similarly, \cite{Sun2022} assessed two additional attribution methods, DeepLIFT \cite{Deeplift} and Integrated Gradients \cite{IntGrad}, for interpreting CNNs trained to predict solar flares. Their study used tracked AR patches within a range of $\pm70^{\circ}$, where the interpretations are confined to central locations.

Furthermore, we presented deep learning-based full-disk flare prediction models trained with limited data in \cite{Pandey2021, Pandey2022} as preliminary studies on their feasibility for a complementary approach to operational forecasting systems. These models collectively showed relative success highlighting a promising direction for the development of more comprehensive full-disk models. However, a common limitation across both models was the lack of explainability. More recently, we presented explainable full-disk flare prediction models \cite{pandeyecml2023, pandeyds2023}, utilizing attribution methods to comprehend the models' effectiveness for near-limb flare events. We provided explanations for a limited set of near-limb flares through a case-based qualitative analysis. In this paper, we extend our approach beyond the case-based qualitative analysis by employing a questionnaire-based evaluation scheme to quantitatively measure post hoc explanations for X-class flares in our dataset. This addition enhances our ability to show the efficacy of our model's predictions for both near-limb and central locations while also verifying its feasibility for operational forecasting systems.

\section{Data}\label{sec:data}

\begin{figure}[tbh!]
\centering
\includegraphics[width=0.98\linewidth ]{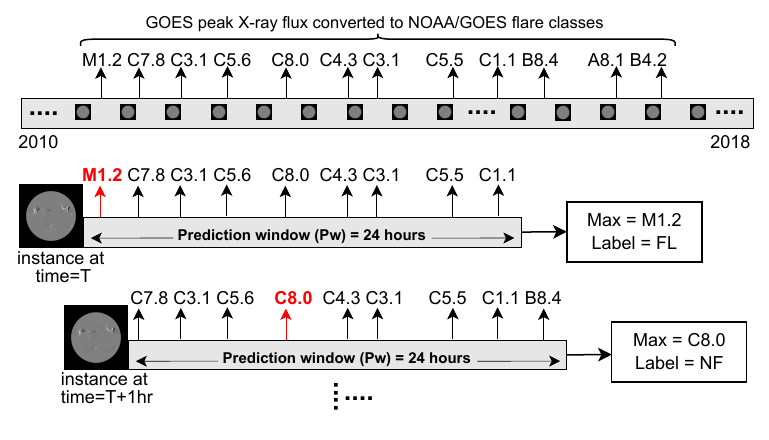}
\caption[]{A visualization for the data labeling procedure of hourly observations of full-disk line-of-sight magnetograms with a prediction window (Pw) of 24 hours. Here, `FL' and `NF' indicate the `Flare' and `No Flare' classes respectively in binary prediction mode ($\geq$M-class flares).}
\label{fig:timeline}
\vspace{-10pt}
\end{figure}

\begin{figure}[tbh!]
\centering
\includegraphics[width=0.95\linewidth ]{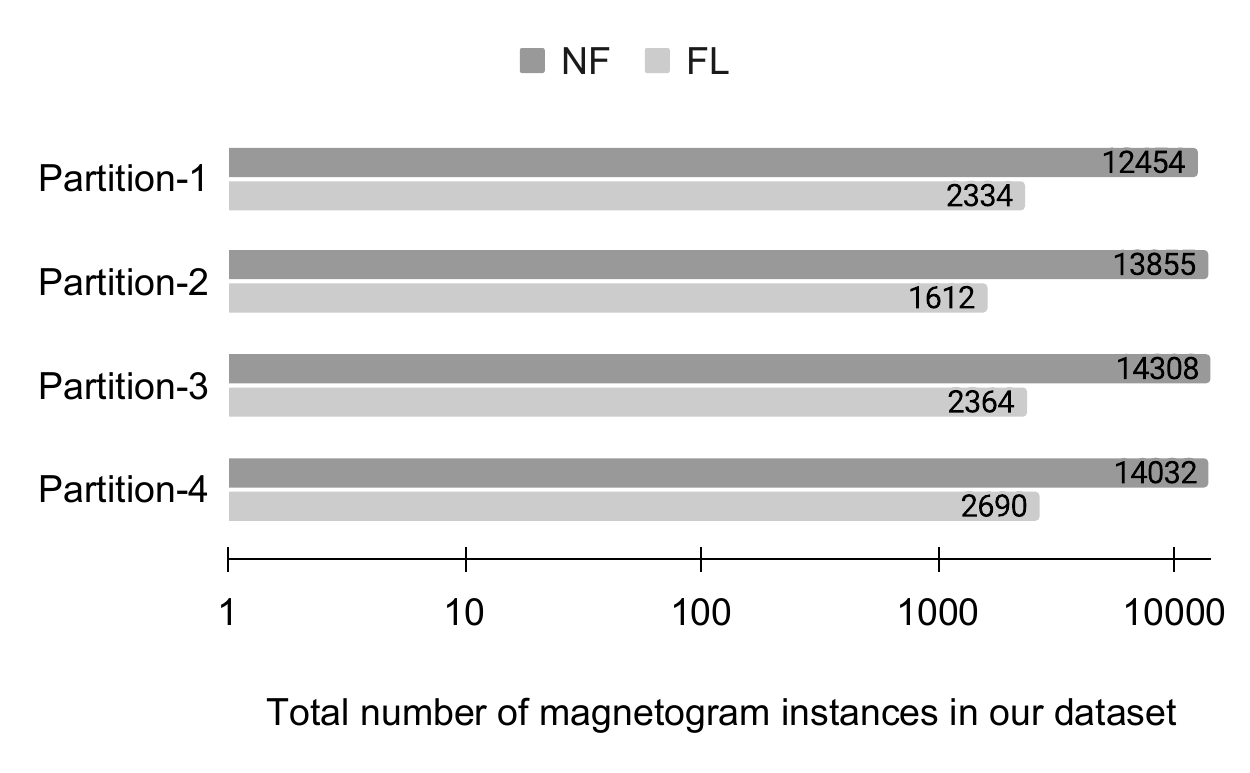}
\caption[]{ Label distribution into four tri-monthly partitions for predicting $\geq$M1.0-class flares. Note that the length of the bar is in logarithmic scale.}
\label{fig:partition}
\vspace{-10pt}
\end{figure}

We used full-disk line-of-sight solar magnetograms obtained from the Helioseismic and Magnetic Imager (HMI) \cite{Schou2011} instrument onboard Solar Dynamics Observatory (SDO) \cite{Pesnell2011} available as compressed JPEG 2000 (JP2) images in near real-time publicly via Helioviewer\cite{Muller2009}. We sampled the magnetogram images every hour of the day, starting at 00:00 and ending at 23:00, from December 2010 to December 2018. We collected a total of 63,649 magnetogram images and labeled them using a 24-hour prediction window based on the maximum peak X-ray flux (converted to NOAA/GOES flare classes), as illustrated in Fig.~\ref{fig:timeline}. To elaborate, for a magnetogram instance at timestamp (T), if the maximum X-ray intensity of a flare was weaker than M within the next 24 hours, we labeled the observation as ``No Flare'' (NF: $<$M), and if it was $\geq$M, we labeled it as ``Flare'' (FL: $\geq$M). This resulted in 54,649 instances for the NF class and 9,000 instances for the FL class.  Finally, we created a non-chronological split of our data into four temporally non-overlapping tri-monthly partitions introduced in \cite{Pandey2021} for our cross-validation experiments, as shown in Fig.~\ref{fig:partition}. Since M- and X-class flares are relatively scarce events, for $\geq$M-class flares the overall distribution of the data becomes highly imbalanced resulting in the FL to NF ratio in our dataset to be $\sim$1:6.


\section{Model and Explanation}\label{sec:modelexp}
\begin{figure}[tbh!]
  \centering
  \includegraphics[width=0.97\linewidth]{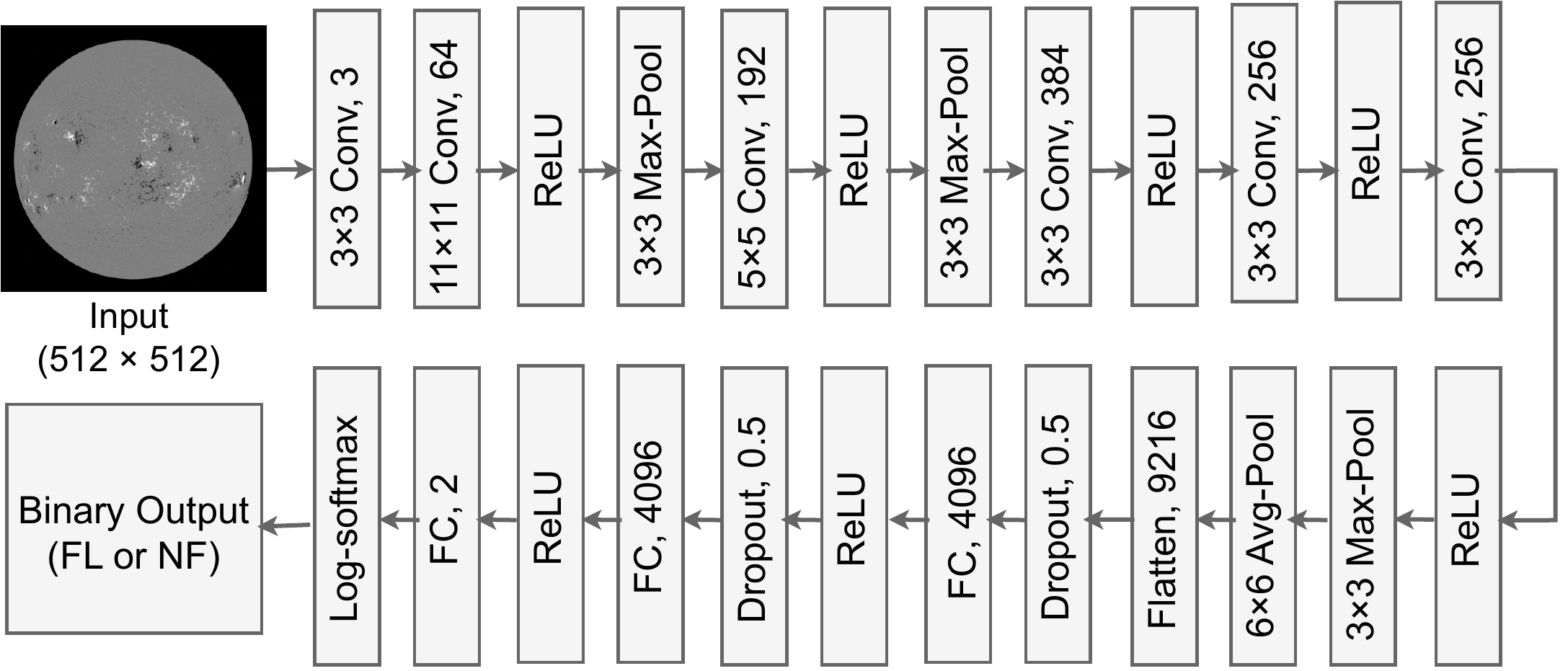}
  \caption{The architecture of our AlexNet-based flare prediction model.}
  \label{fig:arch1}
  \vspace{-5pt}
\end{figure}
In this work, we utilized the AlexNet-based \cite{alex} full-disk flare prediction model in \cite{pandeyds2023}. Our model extends the pre-trained AlexNet model to accommodate 1-channel input magnetogram images by introducing an additional convolutional layer at the beginning of the network that uses a 3$\times$3 kernel, size-1 stride, and outputs three feature maps. This allows the network to take advantage of the pre-trained weights while processing the 1-channel magnetogram images. Moreover, the original AlexNet model is designed to handle 224$\times$224, 3-channel images, whereas our input images are 1-channel, 512$\times$512 images. To address this discrepancy, an adaptive average pooling layer is used after the feature extraction process using the convolutional layer and before the fully-connected layer. This step ensures that the dimensions of the input images are matched to those expected by the pre-trained model. As a result, the model has six convolutional layers, three max pool layers, one adaptive average pool layer, and two fully connected layers. The architecture of the extended AlexNet model used in this study is illustrated in Fig.~\ref{fig:arch1}.

Deep learning models have become increasingly complex, creating difficulties in understanding the intricate data representations they learn. As a result, inconsistencies can arise in the patterns discovered by these models \cite{e23010018}. To address this, attribution methods have been proposed to aid in the interpretation of neural networks' decision-making process. Attribution methods use post hoc attention techniques to generate an attribution vector, or heat map, of the same size as the input, which visualizes how specific input sections contribute to the model's decision without influencing the decision-making process during model training and evaluation. There are mainly two categories of attribution methods: perturbation-based (e.g., Occlusion \cite{Occlusion}) and gradient-based. Perturbation-based methods can lead to inconsistent interpretations due to the creation of Out-of-Distribution data caused by random perturbations \cite{Qiu2022}, while gradient-based methods calculate the gradients of the output with respect to the extracted features or input using backpropagation, enabling attribution scores to be estimated more efficiently and robustly compared to input perturbations \cite{Nielsen2022}. In this study, we employ a gradient-based method namely Guided Grad-CAM \cite{gradcam} to evaluate models and visualize their decisions, identifying the characteristics of magnetogram images that trigger specific decisions to help with operational forecasting under critical conditions.

The Guided Grad-CAM technique \cite{gradcam} is a fusion of the Grad-CAM and guided backpropagation \cite{gbackprop} methods. Grad-CAM is a model-agnostic approach that does not require model retraining, and it employs the class-specific gradient data flowing into the final convolutional layer of a CNN to generate a rough localization map of the image's significant regions. Guided Backpropagation is based on the notion that neurons function as detectors for certain image characteristics. As a result, it computes the output gradient with respect to the input, but it only backpropagates the non-negative gradients through ReLU functions, emphasizing the pixels that are important in the image. Although the attributions generated by Grad-CAM are class-discriminative and localize relevant regions of the image, they do not highlight the pixel-level importance as accurately as guided backpropagation \cite{Chattopadhay2018}. Guided Grad-CAM merges the globally precise details of guided backpropagation with the coarse localization advantages of Grad-CAM and is computed as the element-wise product of guided backpropagation with the upsampled Grad-CAM attributions. We used Captum \cite{kokhlikyan2020captum} library to compute the Guided Grad-CAM attributions and the detailed steps used are enumerated as follows:

\begin{enumerate} [leftmargin=*]
\item Load the pre-trained model and the desired input image $X$ converted to a tensor. Run the forward pass of the model to obtain the final predictions $y$ and the activations $A$ from the last convolution layer.

\item Compute the gradient $\frac{dy_c}{dA_i}$ of the target class score $y_c$ with respect to the activations $A$ at location $i$. This shows how sensitive the class score is to changes in the activations.

\item Calculate the weight $w_i$ for each activation location by averaging the gradients along the channels and spatial dimensions. Compute the Grad-CAM heatmap by performing a weighted sum of the activations followed by Rectified Linear Unit (ReLU) activation: $L_c = \text{ReLU} (\sum_{i} w_i \cdot A_i)$.

\item To compute the guided backpropagation mask:
\begin{enumerate} [leftmargin=0.65cm]
\item Set the gradient of the target class score $y_c$ with respect to the model output as 1, and all other gradients as 0.
\item Perform the backward pass through the network, calculating gradients for all activations and weights. During this process, all negative gradient values are set to 0 at each ReLU activation function.  Hence, the backpropagation process propagates only positive gradients.
\item The output of the guided backpropagation will be a mask $M$, which highlights the pixels in the input that contribute positively to the target class score $y_c$.
\end{enumerate}

\item Combine the Grad-CAM heatmap and guided backpropagation mask by multiplying $L_c$ (upsampled to the input size) element-wise with $M$ to obtain the Guided Grad-CAM heatmap $L^\prime_c = L_c \cdot M$. 
\end{enumerate}

Finally, the Guided Grad-CAM heatmap $L^\prime_c$ is normalized and we can visualize it with or without overlaying on the original image, to identify the important regions contributing to the final prediction.
\section{Experimental Evaluation}\label{sec:expt}
In this section, we describe our experimental settings, including the augmentations used to balance the data, as well as the model implementation and hyperparameter configurations utilized in this study. Furthermore, we present the results and observations drawn from our experiments, with a particular focus on the locations of flares (near-limb flares), which are critical aspects of an operational forecasting system. In addition to reporting on the predictive performance of our models, we also present our findings on post-hoc explanations for X-class flares, which are the most relevant class of flares for Earth-impacting events. To quantify the quality of these explanations, we conducted a questionnaire-based assessment of their reliability, which is also described in this section.
\subsection{Experimental Settings}
The full-disk flare prediction model used in this study is trained using Stochastic Gradient Descent (SGD) as the optimizer and Negative Log-Likelihood (NLL) as the objective function. The model is initialized with pre-trained weights from the AlexNet Model \cite{alex}, then further trained for 40 epochs with a batch size of 64 while employing a dynamic learning rate (initialized at 0.0099 and decreased by 5$\%$). To overcome the issue of class imbalance, we utilized data augmentation and class weights in the loss function. Specifically, we used three augmentation techniques: vertical flipping, horizontal flipping, and rotations ranging from +5$^{\circ}$ to -5$^{\circ}$. We applied augmentation to both classes, tripling the number of augmented samples for the entire FL class and randomly augmenting the NF class. After augmentation, we adjusted the class weights inversely proportional to the class frequencies to penalize misclassifications in the minority class. 

The overall performance of our models is evaluated using two widely-used forecast skills scores: True Skill Statistics (TSS, shown in Eq.~\ref{eq:TSS}) and Heidke Skill Score (HSS, shown in Eq.~\ref{eq:HSS}), derived from the elements of confusion matrix: True Positives (TP), True Negatives (TN), False Positives (FP), and False Negatives (FN). In the context of our solar flare prediction task, the FL class is considered as the positive outcome, while the NF class is negative.

\begin{equation}\label{eq:TSS}
    TSS = \frac{TP}{TP+FN} - \frac{FP}{FP+TN}
\end{equation}

\begin{equation}\label{eq:HSS}
    HSS = 2\times \frac{TP \times TN - FN \times FP}{((P \times (FN + TN) + (TP + FP) \times N))},
\end{equation}
where $N = TN + FP$ and $P = TP + FN$.\\

TSS and HSS values range from -1 to 1, where 1 indicates all correct predictions, -1 represents all incorrect predictions, and 0 represents no skill. In contrast to TSS, HSS is an imbalance-aware metric, and it is common practice to use HSS for the solar flare prediction models due to the high class-imbalance ratio present in the datasets. In solar flare prediction, TSS and HSS are the preferred evaluation metrics over commonly used ones in image classification, such as accuracy. This preference is because they offer a more thorough and dependable way to assess how well predictions work, especially when dealing with imbalanced class distributions. Lastly, we report the subclass and overall recall (shown in Eq.~\ref{eq:rec}) for FL class instances (M-class and X-class) to demonstrate the prediction sensitivity in central and near-limb regions. 
\begin{equation}\label{eq:rec}
    Recall = \frac{TP}{TP+FN}
\end{equation}
To reproduce this work, the source code and experimental results can be accessed from our open-source repository \cite{sourcecode}. 
\subsection{Model Evaluation}
The aggregated results from the 4-fold cross-validation experiments are as follows: Our models have on average TSS$\sim$0.51 and HSS$\sim$0.38, which improves over the performance of \cite{Pandey2021} by $\sim$4\% in terms of TSS (reported $\sim$0.47) and $\sim$3\% in terms of HSS (reported $\sim$0.35). In addition, on evaluating our results for correctly predicted (TP) and missed flare (FN) counts for class-specific flares (X-class and M-class) in central locations (within $\pm$70$^{\circ}$) and near-limb locations (beyond $\pm$70$^{\circ}$) of the Sun as shown in Table \ref{table:comp}, we observe that our models made correct predictions for $\sim$95\% of the X-class flares and $\sim$73\% of the M flares in central locations. Similarly, our models show a compelling performance for flares appearing on near-limb locations of the Sun, where $\sim$74\% of the X-class and $\sim$50\%  of the M-class flares are predicted correctly\footnote{It is important to note that there are certain flares that may not be predicted at all with current observational capabilities, particularly when the active region responsible for the flare is not visible or only partially visible on the East limb up to 24 hours prior to the occurrence of the flare.}. This is important because, to our knowledge, the prediction of near-limb flares is often overlooked. More false positives in M-class are expected because of the model's inability to distinguish bordering class (C4+ to C9.9) flares from $\geq$M1.0-class flares, which we have observed empirically in our prior work as well. Overall, we observed that $\sim$90\% and $\sim$66\% of the X-class and M-class flares, respectively, are predicted correctly by our models.

\begin{figure*}[tbh!]
\centering
\includegraphics[width=0.85\linewidth ]{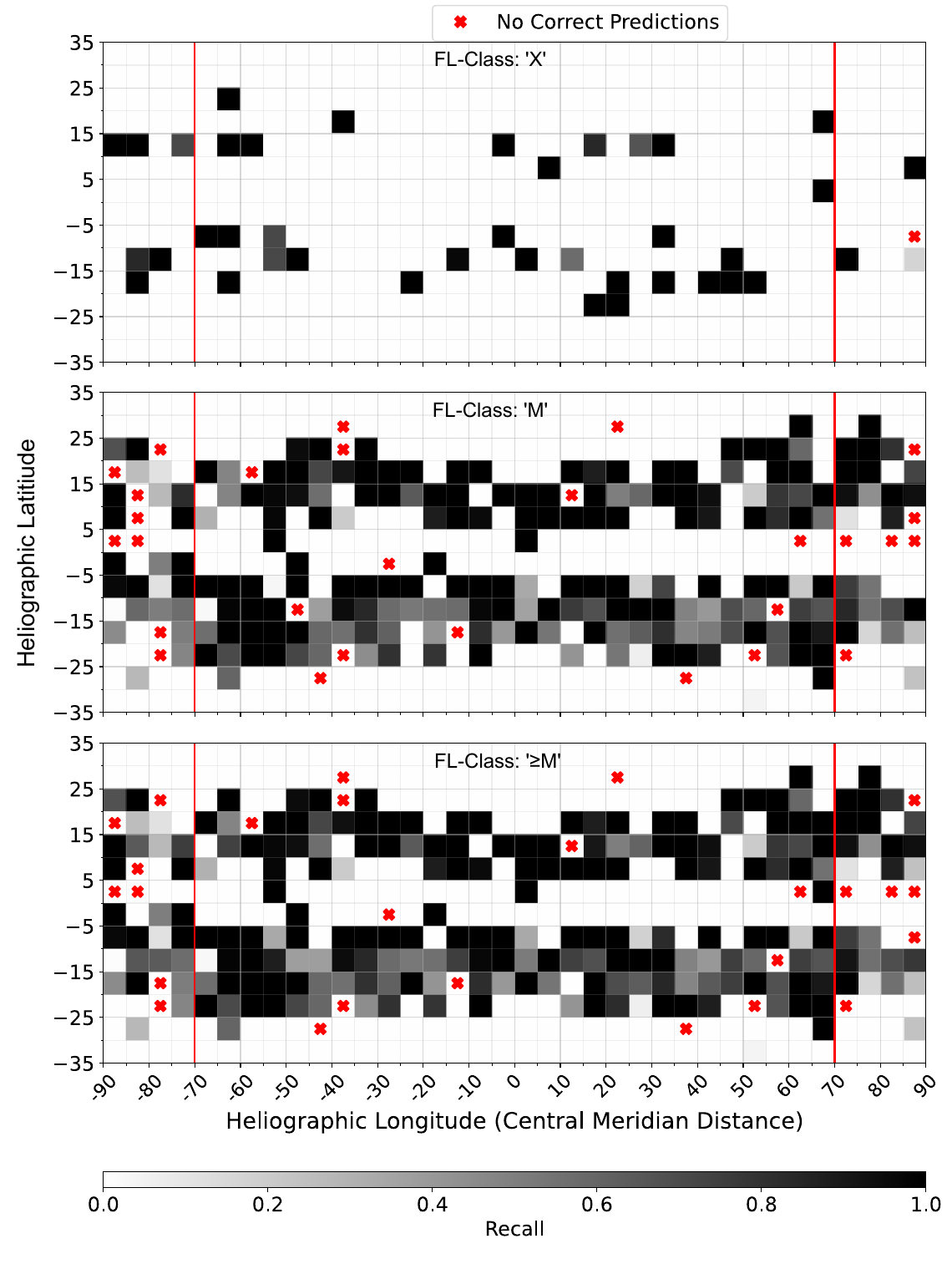}
\vspace{-10pt}
\caption[]{A heatmap to assess the performance of our models in terms of recall for individual FL-Class (X- and M-class flares) and when combined ($\geq$M-class flares) binned into 5$^{\circ}$ $\times$ 5$^{\circ}$ flare locations used as the label. The flare events including and beyond $\pm$70$^{\circ}$ longitude (separated by a vertical red line) are represented as near-limb events.  Note: (i) Red cross in white grids represents locations with zero correct predictions while white cells without red cross represent unavailable instances. (ii) These results are aggregated from the validation sets of 4-folds.}
\label{fig:histogram}
\vspace{-10pt}
\end{figure*}

\begin{table}[tbh!]
\setlength{\tabcolsep}{3pt}
\renewcommand{\arraystretch}{1.5}
\caption{Counts of correctly (TP) and incorrectly (FN) classified X- and M-class flares in central ($|longitude|$$\leq\pm70^{\circ}$) and near-limb locations. The recall across different location groups is also presented. Counts are aggregated across folds.}
\begin{center}
 \begin{tabular}{r c c c c c c}
\hline
 & 
\multicolumn{3}{c}{Within $\pm$70$^{\circ}$} 
&                                            
\multicolumn{3}{c}{Beyond $\pm$70$^{\circ}$}\\
Flare-Class & TP  & FN  & Recall  & TP   &FN & Recall \\
\hline
X-Class  &  637  & 31  & 0.95 & 157 & 55 & 0.74\\

M-Class &  4229 & 1601  & 0.73 & 1143   &1147 & 0.50\\

Total (X\&M) & 4866 & 1632 &0.75& 1300 & 1202 & 0.52\\ 
\hline
\end{tabular}
\end{center}
\label{table:comp}
\vspace{-10pt}
\end{table}



We evaluated the effectiveness of our models both quantitatively and qualitatively by spatially analyzing their performance with respect to the locations of M- and X-class flares responsible for the labels. To conduct our analysis, we spatially binned the responsible flares based on their location in the Heliographic Stonyhurst (HGS) coordinate system, where each bin represents a 5$^{\circ}$ by 5$^{\circ}$ spatial cell in terms of latitude and longitude. Then, we analyzed the predictions of our models in the validation set from the 4-fold cross-validation experiments to determine whether the instances were correctly or incorrectly predicted. We calculated the recall separately for M-class, X-class, and both M- and X-class flares in each spatial cell to assess the models' sensitivity at a fine-grained level. The heatmaps that illustrate the spatial distribution of recall scores for our models can be found in Fig.~\ref{fig:histogram}. This allowed us to pinpoint the locations where our models were more effective in making accurate predictions and vice versa.

Our research findings indicate that our models are capable of effectively predicting even near-limb flares, particularly for X-class flares in both central and near-limb locations. However, we observed a higher number of false negatives in the near-limb areas for M-class flares. This represents a new capability in space weather forecasting, as we can now predict flares in regions where the magnetic field is distorted and pinpoint the locations of relevant active regions that are more likely to flare, which is critical for operational forecasting methods.

\subsection{Evaluating Post hoc Explanations}

\begin{figure*}[tbh!]
\centering
\begin{tabular}{c}
\includegraphics[width=0.99\linewidth]{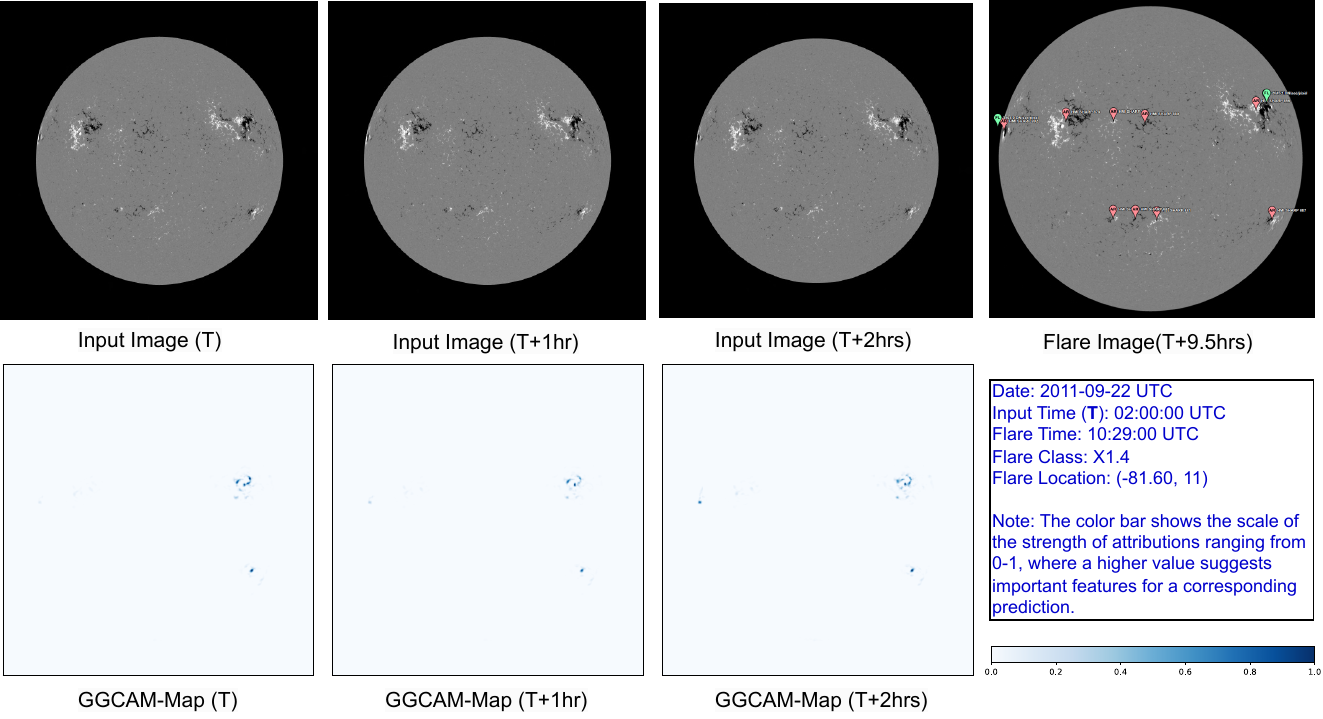} \\
(a) \\
\\
\end{tabular}
\begin{tabular}{c}
\includegraphics[width=0.99\linewidth]{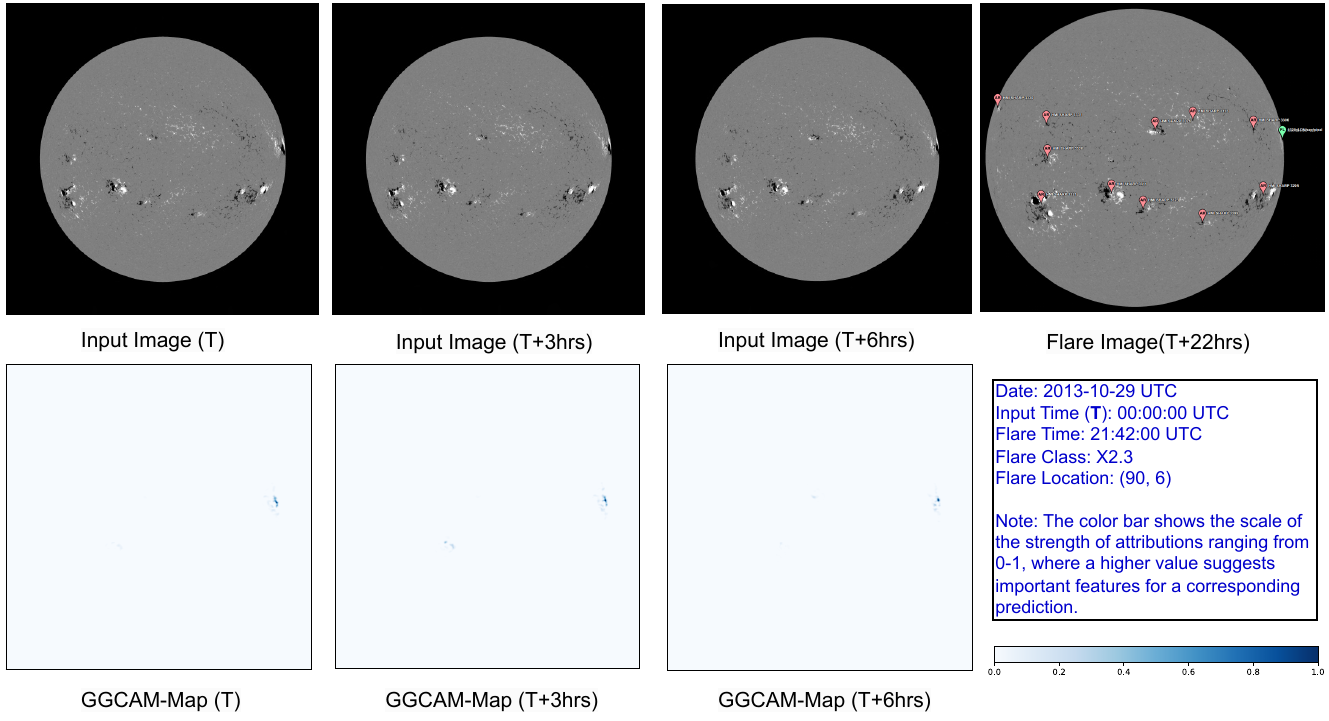} \\
(b)\\
\end{tabular}
\caption{A visual explanation of two correctly predicted near-limb X-class flares for East and West limbs, depicted in subfigures (a) and (b), respectively. The top row displays a series of three full-disk magnetogram input images, with a temporal progression of i=\{0, 1, 2\} hours for (a) and i=\{0, 3, 6\} hours for (b), starting from an initial timestamp T (different for each subfigure). These images are followed by a magnetogram image at approximately T+k hours, featuring overlays of detected active regions at the time of the flare (k$\sim$8.5 hours in (a) and k$\sim$22 hours in (b), denoted as Flare Image). Green flags indicate the flare location, while red flags represent all active regions in the magnetogram. The bottom row exhibits three activation maps generated using Guided Grad-CAM, corresponding to the input images and following the same temporal progression respective to subfigures (a) and (b).}
\label{fig:vis}
\vspace{-10pt}
\end{figure*}
\begin{table}[bt!]
\setlength{\tabcolsep}{3pt}
\renewcommand{\arraystretch}{1.5}
\caption{Total counts of Guided Grad-CAM (True when activated and False otherwise) that corresponds to the flare location (FLOC: location of maximum peak X-ray flux used to label each magnetogram instances) for correctly (TP) and incorrectly (FN) classified  X-class flares in central ($|longitude|$$\leq\pm70^{\circ}$) and near-limb locations.}
\begin{center}
 \begin{tabular}{r c c c c c c}
\hline
 & 
\multicolumn{2}{c}{Within $\pm$70$^{\circ}$} 
& 
\multicolumn{2}{c}{Beyond $\pm$70$^{\circ}$} 
&                                            
\multicolumn{2}{c}{Full-disk (Total)}\\
Prediction & True  & False  & True  & False   &True & False \\
\hline
TP  &  635  & 2  & 149 & 8 & 784 & 10\\

FN &  31 & 0  & 55 & 0   & 86 & 0\\ 
\hline
\end{tabular}
\end{center}
\label{table:exp1}
\vspace{-15pt}
\end{table}

\begin{table*}[tbh!]
\setlength{\tabcolsep}{5pt}
\renewcommand{\arraystretch}{1.5}
\caption{Total counts of Guided Grad-CAM that corresponds to the responsible flare location (FLOC), other X-flares location (beside maximum intensity flare event, i.e., FL-class) denoted as XLOC, other M-flares location denoted as MLOC, and other C-flares location denoted as CLOC, which coexists within the prediction window of 24 hours. The count is presented for correctly (TP) and incorrectly (FN) classified  X-class flares (FL) in central ($|longitude|$$\leq\pm70^{\circ}$) and near-limb locations.}

\begin{center}
 \begin{tabular}{|r c c| c c| c c| c c|}
\hline
\multicolumn{3}{|c|}{\textbf{When FL-location is activated in GGCAM (FLOC=T)}}
 & 
\multicolumn{2}{c|}{Within $\pm$70$^{\circ}$} 
& 
\multicolumn{2}{c|}{Beyond $\pm$70$^{\circ}$} 
&                                            
\multicolumn{2}{c|}{Full-disk (Total)}\\
\hline
Coexisting X-Flare  & Coexisting M-flare  & Coexisting C-Flare   &  &   &  &   &  &  \\
Locations (XLOC) & Locations (MLOC) & Locations (CLOC) & TP  & FN  & TP  & FN & TP & FN \\
\hline
T& T &T  &  8  & 0  & 4 & 0 & 12 & 0\\

T & N/A & T  &  11  & 0  & 0 & 0 & 11 & 0\\

N/A& T & N/A  &  25  & 4  & 0 & 0 & 25 & 4\\

N/A & T & T  &  398  & 6  & 47 & 17 & 445 & 23\\

N/A& N/A &T  &  186  & 21  & 98 & 38 & 284 & 59\\

N/A& F& T  &  7  & 0  & 0 & 0 & 7 & 0\\
\hline
\multicolumn{3}{|c|}{\textbf{When FL-location is not activated in GGCAM (FLOC=F)}}
 & 
\multicolumn{2}{c|}{} 
& 
\multicolumn{2}{c|}{} 
&                                            
\multicolumn{2}{c|}{}\\
\hline

N/A& T& T  &  0  & 0  & 6 & 0 & 6 & 0\\

N/A& N/A& T  &  0  & 0  & 1 & 0 & 1 & 0\\

N/A& F& T  &  2  & 0  & 1 & 0 & 3 & 0\\
\hline
\end{tabular}
\end{center}
\label{table:exp2}
\vspace{-15pt}
\end{table*}
In this section, we present a comprehensive analysis of the explanations generated using the Guided Grad-CAM attribution method for X-class flares. We focus on X-class flares because they are the most significant contributors to Earth-impacting extreme space weather events, such as the Halloween and Bastille Day Superstorms \cite{Tsurutani2006}. Our analysis encompasses both qualitative and quantitative evaluations of the generated explanations, providing a more detailed understanding of the attribution method's effectiveness in predicting X-class flares. By examining the results closely, we aim to improve our ability to forecast these powerful flares and mitigate their potential impacts.

For qualitative analytics, we evaluate the visual explanations with their temporal progression and interpret them by showcasing two examples of correctly predicted near-limb flares on both the East and West limbs of the Sun. Firstly, we present the explanation of our model for an East-limb X1.4-class (note that East and West are reversed in solar coordinates) flare observed on 2011-09-22 at 10:29:00 UTC with visualizations generated using Guided Grad-CAM, denoted as GGCAM-MAP in Fig.~\ref{fig:vis} (a). For this, we used three input images, each separated by an hour starting at 2011-09-22 02:00:00 UTC (denoted as T), until the sunspot for the corresponding flare starts to become more visible in the magnetogram image. With temporal progression, we observed that as soon as a region becomes visible, the pixels covering the AR on the East limb are activated as shown in Fig.~\ref{fig:vis} (a), in fact, the model focuses on specific ARs including the relatively smaller area AR (in terms of pixels) on the East limb, even though other ARs are present in the magnetogram image. The visualization of attribution maps suggests that for this particular prediction, although only partially visible due to rotation and projection effects, the region responsible for the flare event is attributed to be important and hence contributes to the consequent decision.  

Similarly, we analyze another case of correctly predicted near-limb flare (West-limb) of the Sun. For this, we provide a case of X2.3-class flare observed on 2013-10-29 at 21:42:00 UTC. For this, we again used three input images, each separated by three hours starting on 2013-10-29 at 00:00:00 UTC (denoted as T) shown in Fig.~\ref{fig:vis} (b). We observe that for the duration of six hours (until 16 hours prior to the flare event with three input images), the model's activation for AR on the West-limb remains intact and it remains so as long as the AR corresponding to the flare is visible on the full-disk, indicating it is an important region for corresponding correct prediction. While there are several larger ARs present in the magnetogram, it focuses on specific ARs including the relatively smaller AR on the West limb. This shows that our models were able to learn shape-based features from compressed images and use them effectively for predicting flares, even in the presence of severe projection effects.

In addition, we conducted a rigorous quantitative evaluation of explanations for X-class flares, which comprised 880 instances in our dataset. We adopted a human-centered approach, as mentioned in \cite{explainai}, and designed a set of questionnaires addressing each instance to validate the explanations. The questionnaires were structured as follows:

\begin{enumerate}[leftmargin=*]
\item Determine the activation of the location of the responsible X-class flare, denoted as FL, with maximum peak X-ray flux used to label the image. Assess the activation using a binary response format, with True (T) indicating activation and False (F) indicating non-activation. The location of FL is denoted as FLOC=\{T, F\}.

\item Determine the activation of any other X-class flare locations besides the maximum flare location (FLOC) used to label the instance (i.e., FL) that coexisted within the 24-hour prediction window. Use a ternary response format, with `T' indicating activation, `F' indicating non-activation, and Not Applicable (N/A) indicating the absence of other X-class flares in the prediction window. This is denoted as XLOC=\{T, F, N/A\}.

\item Determine the activation of any M-class flare locations that coexisted with FL within the 24-hour prediction window. Use a ternary response format, with `T' indicating activation, `F' indicating non-activation, and `N/A' indicating the absence of M-class flares in the prediction window. This is denoted as MLOC=\{T, F, N/A\}.

\item Determine the activation of any C-class flare locations that coexisted with FL within the 24-hour prediction window. Use a ternary response format, with `T' indicating activation, `F' indicating non-activation, and `N/A' indicating the absence of C-class flares in the prediction window. This is denoted as CLOC=\{T, F, N/A\}.
\end{enumerate}

After examining the outcomes of question (1), which evaluated the activation of location for flare with maximum peak X-ray flux (FLOC) for X-class flares in our dataset, we found that FLOC was activated in 870 out of 880 instances, as presented in Table~\ref{table:exp1}. We subsequently conducted a separate evaluation of the outcomes for instances predicted correctly (TP) and incorrectly (FN) in both central and near-limb locations\footnote{Note that the central and near-limb locations referred to are only for the corresponding FL location (FLOC), and the evaluations are presented irrespective of the locations of other coexisting X, M, and C-class flares. Furthermore, the activations maps for incorrectly predicted instances (FN) are also generated and evaluated for positive class labels.}. Upon analyzing the results, we discovered that in every false negative (FN) instance, the model was able to locate the location of the flare but could not classify it accurately. Additionally, we observed that there were 10 instances ($\sim$1.1\% of total X-class flares), of which 8 were located in near-limb regions, while the remaining 2 were located in central locations, that were predicted correctly. However, the FLOC was not activated because other active regions were also activated, and those regions produced one or more M- and C-class flares. The location of the responsible flares was partially or completely invisible while in the near-limb region. Nevertheless, the model was able to predict these instances correctly.

In addition to question (1), we also incorporated the results of questions (2), (3), and (4) to examine the impact of other flares on the corresponding predictions, as presented in Table~\ref{table:exp2}. Notably, in instances where the FLOC is activated, we observed that only 23 instances had all other flare locations (i.e., XLOC, MLOC, and CLOC, if not N/A) activated. Furthermore, we found that all of the CLOC (if not N/A) were activated, even when the MLOC and FLOC were not activated. Overall, the majority of instances ($\sim$53\%) showed activations for both MLOC and CLOC (if they coexist) along with FLOC. Our findings here point to a well-known phenomenon that flares productive active regions emit flares and eruptions in a clustered fashion \cite{Toriumi2019, Campi2019}. Moreover, we observed that, for almost 39\% of the instances, where M-class locations were absent, only CLOC along with FLOC were activated. This may be due to the higher rates of occurrence of C-class flares, which constitute approximately 40\% (25,150 instances) of the entire dataset and the model struggles to distinguish them from the FL-class as observed in our prior work \cite{Pandey2022}. Despite this, we observed the model's adept ability to accurately identify relevant locations for the corresponding prediction, even for near-limb events, which emphasizes the utility of the full-disk model as a valuable contribution to operationalization efforts. Furthermore, it's important to note that the explanations show our model's ability to selectively highlight specific ARs that are more likely to initiate flares. This aspect is crucial for localizing flare-productive ARs, which is of utmost importance when performing long-term SEP event prediction \cite{ji2021modular}.

\section{Conclusion and Future Work}\label{sec:conc}
In this work, we used a gradient-based attribution method, namely, Guided Grad-CAM to interpret the predictions of our binary ( $\geq$M1.0-class) full-disk flare prediction model. We addressed the highly overlooked problem of flares appearing in near-limb locations of the Sun, and our evaluation shows a compelling performance of a full-disk model for such events. Furthermore, we evaluated our model's predictions with visual explanations and showed that the model's decisions are primarily based on characteristics corresponding to the relevant active regions in the magnetogram instances. As an extension, we plan to develop an automated method of evaluating explanations and incorporate the M-class flares as well to make this study more comprehensive. Furthermore, at this point, our models consider only the point-in-time spatial raster features in our data, and we intend to widen this work toward building models that can capture the temporal evolution of these features to improve predictive performance.

\section*{Acknowledgements}
This project is supported in part under two NSF awards \#2104004 and \#1931555, jointly by the Office of Advanced Cyberinfrastructure within the Directorate for Computer and Information Science and Engineering, the Division of Astronomical Sciences within the Directorate for Mathematical and Physical Sciences, and the Solar Terrestrial Physics Program and the Division of Integrative and Collaborative Education and Research within the Directorate for Geosciences. This work is also partially supported by the National Aeronautics and Space Administration (NASA) grant award \#80NSSC22K0272. Data used in this study is a courtesy of NASA/SDO and the AIA, EVE, and HMI science teams and NOAA National Geophysical Data Center (NGDC).
\bibliographystyle{IEEEtran}
\bibliography{bibwodoi}

\begin{thebibliography}{10}
\providecommand{\url}[1]{#1}
\csname url@samestyle\endcsname
\providecommand{\newblock}{\relax}
\providecommand{\bibinfo}[2]{#2}
\providecommand{\BIBentrySTDinterwordspacing}{\spaceskip=0pt\relax}
\providecommand{\BIBentryALTinterwordstretchfactor}{4}
\providecommand{\BIBentryALTinterwordspacing}{\spaceskip=\fontdimen2\font plus
\BIBentryALTinterwordstretchfactor\fontdimen3\font minus
  \fontdimen4\font\relax}
\providecommand{\BIBforeignlanguage}[2]{{%
\expandafter\ifx\csname l@#1\endcsname\relax
\typeout{** WARNING: IEEEtran.bst: No hyphenation pattern has been}%
\typeout{** loaded for the language `#1'. Using the pattern for}%
\typeout{** the default language instead.}%
\else
\language=\csname l@#1\endcsname
\fi
#2}}
\providecommand{\BIBdecl}{\relax}
\BIBdecl

\bibitem{spaceweather}
L.~Fletcher, B.~R. Dennis, H.~S. Hudson, S.~Krucker, K.~Phillips, A.~Veronig,
  M.~Battaglia, L.~Bone, A.~Caspi, Q.~Chen, P.~Gallagher, P.~T. Grigis, H.~Ji,
  W.~Liu, R.~O. Milligan, and M.~Temmer, ``An observational overview of solar
  flares,'' \emph{Space Science Reviews}, vol. 159, no. 1-4, pp. 19--106, Aug.
  2011.

\bibitem{Yasyukevich2018}
Y.~Yasyukevich, E.~Astafyeva, A.~Padokhin, V.~Ivanova, S.~Syrovatskii, and
  A.~Podlesnyi, ``The 6 september 2017 x-class solar flares and their impacts
  on the ionosphere, {GNSS}, and {HF} radio wave propagation,'' \emph{Space
  Weather}, vol.~16, no.~8, pp. 1013--1027, Aug. 2018.

\bibitem{Pandey2022f}
\BIBentryALTinterwordspacing
C.~Pandey, A.~Ji, R.~A. Angryk, M.~K. Georgoulis, and B.~Aydin, ``Towards
  coupling full-disk and active region-based flare prediction for operational
  space weather forecasting,'' \emph{Frontiers in Astronomy and Space
  Sciences}, vol.~9, Aug. 2022. [Online]. Available:
  \url{https://doi.org/10.3389/fspas.2022.897301}
\BIBentrySTDinterwordspacing

\bibitem{Falconer2016}
D.~A. Falconer, S.~K. Tiwari, R.~L. Moore, and I.~Khazanov, ``A new method to
  quantify and reduce the net projection error in whole-solar-active-region
  parameters measured from vector magnetograms,'' \emph{The Astrophysical
  Journal}, vol. 833, no.~2, p. L31, Dec. 2016.

\bibitem{Huang2018}
X.~Huang, H.~Wang, L.~Xu, J.~Liu, R.~Li, and X.~Dai, ``Deep learning based
  solar flare forecasting model. i. results for line-of-sight magnetograms,''
  \emph{The Astrophysical Journal}, vol. 856, no.~1, p.~7, Mar. 2018.

\bibitem{Ji2020}
A.~Ji, B.~Aydin, M.~K. Georgoulis, and R.~Angryk, ``All-clear flare prediction
  using interval-based time series classifiers,'' in \emph{2020 {IEEE}
  International Conference on Big Data (Big Data)}.\hskip 1em plus 0.5em minus
  0.4em\relax {IEEE}, Dec. 2020, pp. 4218--4225.

\bibitem{Hoeksema2014}
J.~T. Hoeksema, Y.~Liu, K.~Hayashi, X.~Sun, J.~Schou, S.~Couvidat, A.~Norton,
  M.~Bobra, R.~Centeno, K.~D. Leka, G.~Barnes, and M.~Turmon, ``The
  helioseismic and magnetic imager ({HMI}) vector magnetic field pipeline:
  Overview and performance,'' \emph{Solar Physics}, vol. 289, no.~9, pp.
  3483--3530, Mar. 2014.

\bibitem{Korss2014}
M.~B. Kors{\'{o}}s, T.~Baranyi, and A.~Ludm{\'{a}}ny, ``{PRE}-{FLARE}
  {DYNAMICS} {OF} {SUNSPOT} {GROUPS},'' \emph{The Astrophysical Journal}, vol.
  789, no.~2, p. 107, Jun. 2014.

\bibitem{McCloskey2016}
A.~E. McCloskey, P.~T. Gallagher, and D.~S. Bloomfield, ``Flaring rates and the
  evolution of sunspot group {McIntosh} classifications,'' \emph{Solar
  Physics}, vol. 291, no.~6, pp. 1711--1738, Jun. 2016.

\bibitem{Deshmukh2020}
V.~Deshmukh, T.~E. Berger, E.~Bradley, and J.~D. Meiss, ``Leveraging the
  mathematics of shape for solar magnetic eruption prediction,'' \emph{Journal
  of Space Weather and Space Climate}, vol.~10, p.~13, 2020.

\bibitem{Ji2023}
A.~Ji, X.~Cai, N.~Khasayeva, M.~K. Georgoulis, P.~C. Martens, R.~A. Angryk, and
  B.~Aydin, ``A systematic magnetic polarity inversion line data set from
  {SDO}/{HMI} magnetograms,'' \emph{The Astrophysical Journal Supplement
  Series}, vol. 265, no.~1, p.~28, Mar. 2023.

\bibitem{Nishizuka_2017}
N.~Nishizuka, K.~Sugiura, Y.~Kubo, M.~Den, S.~Watari, and M.~Ishii, ``Solar
  flare prediction model with three machine-learning algorithms using
  ultraviolet brightening and vector magnetograms,'' \emph{The Astrophysical
  Journal}, vol. 835, no.~2, p. 156, jan 2017.

\bibitem{Nishizuka2018}
N.~Nishizuka, K.~Sugiura, Y.~Kubo, M.~Den, and M.~Ishii, ``Deep flare net
  ({DeFN}) model for solar flare prediction,'' \emph{The Astrophysical
  Journal}, vol. 858, no.~2, p. 113, May 2018.

\bibitem{Li2020}
X.~Li, Y.~Zheng, X.~Wang, and L.~Wang, ``Predicting solar flares using a novel
  deep convolutional neural network,'' \emph{The Astrophysical Journal}, vol.
  891, no.~1, p.~10, Feb. 2020.

\bibitem{Pandey2021}
\BIBentryALTinterwordspacing
C.~Pandey, R.~A. Angryk, and B.~Aydin, ``Solar flare forecasting with deep
  neural networks using compressed full-disk {HMI} magnetograms,'' in
  \emph{2021 {IEEE} International Conference on Big Data (Big Data)}.\hskip 1em
  plus 0.5em minus 0.4em\relax {IEEE}, Dec. 2021, pp. 1725--1730. [Online].
  Available: \url{https://doi.org/10.1109/bigdata52589.2021.9671322}
\BIBentrySTDinterwordspacing

\bibitem{Pandey2022}
\BIBentryALTinterwordspacing
C.Pandey, R.~A. Angryk, and B.~Aydin, ``Deep neural networks based solar flare
  prediction using compressed full-disk line-of-sight magnetograms,'' in
  \emph{Information Management and Big Data}.\hskip 1em plus 0.5em minus
  0.4em\relax Springer International Publishing, 2022, pp. 380--396. [Online].
  Available: \url{https://doi.org/10.1007/978-3-031-04447-2\_26}
\BIBentrySTDinterwordspacing

\bibitem{Whitman2022}
\BIBentryALTinterwordspacing
K.~Whitman, R.~Egeland, I.~G. Richardson, C.~Allison, P.~Quinn, J.~Barzilla,
  I.~Kitiashvili, V.~Sadykov, H.~M. Bain, M.~Dierckxsens, M.~L. Mays,
  T.~Tadesse, K.~T. Lee, E.~Semones, J.~G. Luhmann, M.~N{\'{u}}{\~{n}}ez, S.~M.
  White, S.~W. Kahler, A.~G. Ling, D.~F. Smart, M.~A. Shea, V.~Tenishev, S.~F.
  Boubrahimi, B.~Aydin, P.~Martens, R.~Angryk, M.~S. Marsh, S.~Dalla,
  N.~Crosby, N.~A. Schwadron, K.~Kozarev, M.~Gorby, M.~A. Young, M.~Laurenza,
  E.~W. Cliver, T.~Alberti, M.~Stumpo, S.~Benella, A.~Papaioannou,
  A.~Anastasiadis, I.~Sandberg, M.~K. Georgoulis, A.~Ji, D.~Kempton, C.~Pandey,
  G.~Li, J.~Hu, G.~P. Zank, E.~Lavasa, G.~Giannopoulos, D.~Falconer, Y.~Kadadi,
  I.~Fernandes, M.~A. Dayeh, A.~Mu{\~{n}}oz-Jaramillo, S.~Chatterjee, K.~D.
  Moreland, I.~V. Sokolov, I.~I. Roussev, A.~Taktakishvili, F.~Effenberger,
  T.~Gombosi, Z.~Huang, L.~Zhao, N.~Wijsen, A.~Aran, S.~Poedts,
  A.~Kouloumvakos, M.~Paassilta, R.~Vainio, A.~Belov, E.~A. Eroshenko, M.~A.
  Abunina, A.~A. Abunin, C.~C. Balch, O.~Malandraki, M.~Karavolos, B.~Heber,
  J.~Labrenz, P.~K\"{u}hl, A.~G. Kosovichev, V.~Oria, G.~M. Nita,
  E.~Illarionov, P.~M. O'Keefe, Y.~Jiang, S.~H. Fereira, A.~Ali, E.~Paouris,
  S.~Aminalragia-Giamini, P.~Jiggens, M.~Jin, C.~O. Lee, E.~Palmerio, A.~Bruno,
  S.~Kasapis, X.~Wang, Y.~Chen, B.~Sanahuja, D.~Lario, C.~Jacobs, D.~T.
  Strauss, R.~Steyn, J.~van~den Berg, B.~Swalwell, C.~Waterfall, M.~Nedal,
  R.~Miteva, M.~Dechev, P.~Zucca, A.~Engell, B.~Maze, H.~Farmer, T.~Kerber,
  B.~Barnett, J.~Loomis, N.~Grey, B.~J. Thompson, J.~A. Linker, R.~M. Caplan,
  C.~Downs, T.~T\"{o}r\"{o}k, R.~Lionello, V.~Titov, M.~Zhang, and
  P.~Hosseinzadeh, ``Review of solar energetic particle models,''
  \emph{Advances in Space Research}, Aug. 2022. [Online]. Available:
  \url{https://doi.org/10.1016/j.asr.2022.08.006}
\BIBentrySTDinterwordspacing

\bibitem{Hong2023}
J.~Hong, A.~Ji, C.~Pandey, and B.~Aydin, ``Beyond traditional flare
  forecasting: A data-driven labeling approach for~high-fidelity predictions,''
  in \emph{Big Data Analytics and Knowledge Discovery}.\hskip 1em plus 0.5em
  minus 0.4em\relax Springer Nature Switzerland, 2023, pp. 380--385.

\bibitem{pandeyds2023}
\BIBentryALTinterwordspacing
C.~Pandey, R.~A. Angryk, M.~K. Georgoulis, and B.~Aydin, ``Explainable deep
  learning-based solar flare prediction with post hoc attention for operational
  forecasting,'' 2023. [Online]. Available:
  \url{https://arxiv.org/abs/2308.02682}
\BIBentrySTDinterwordspacing

\bibitem{explainai}
G.~Vilone and L.~Longo, ``Notions of explainability and evaluation approaches
  for explainable artificial intelligence,'' \emph{Inf. Fusion}, vol.~76,
  no.~C, p. 89–106, dec 2021.

\bibitem{Crown2012}
M.~D. Crown, ``Validation of the {NOAA} space weather prediction
  center{\textquotesingle}s solar flare forecasting look-up table and
  forecaster-issued probabilities,'' \emph{Space Weather}, vol.~10, no.~6, pp.
  n/a--n/a, Jun. 2012.

\bibitem{Devos2014}
A.~Devos, C.~Verbeeck, and E.~Robbrecht, ``Verification of space weather
  forecasting at the regional warning center in belgium,'' \emph{Journal of
  Space Weather and Space Climate}, vol.~4, p. A29, 2014.

\bibitem{Lee2012}
K.~Lee, Y.-J. Moon, J.-Y. Lee, K.-S. Lee, and H.~Na, ``Solar flare occurrence
  rate and probability in terms of the sunspot classification supplemented with
  sunspot area and its changes,'' \emph{Solar Physics}, vol. 281, no.~2, pp.
  639--650, Sep. 2012.

\bibitem{Leka2018}
K.~Leka, G.~Barnes, and E.~Wagner, ``The {NWRA} classification infrastructure:
  description and extension to the discriminant analysis flare forecasting
  system ({DAFFS}),'' \emph{Journal of Space Weather and Space Climate},
  vol.~8, p. A25, 2018.

\bibitem{Kusano2020}
K.~Kusano, T.~Iju, Y.~Bamba, and S.~Inoue, ``A physics-based method that can
  predict imminent large solar flares,'' \emph{Science}, vol. 369, no. 6503,
  pp. 587--591, Jul. 2020.

\bibitem{Korss2020}
M.~B. Kors{\'{o}}s, M.~K. Georgoulis, N.~Gyenge, S.~K. Bisoi, S.~Yu, S.~Poedts,
  C.~J. Nelson, J.~Liu, Y.~Yan, and R.~Erd{\'{e}}lyi, ``Solar flare prediction
  using magnetic field diagnostics above the photosphere,'' \emph{The
  Astrophysical Journal}, vol. 896, no.~2, p. 119, Jun. 2020.

\bibitem{Bobra2015}
M.~G. Bobra and S.~Couvidat, ``Solar flare prediction
  using$\leq${SDO}$\leq$/{HMI} {VECTOR} {MAGNETIC} {FIELD} {DATA} {WITH} a
  {MACHINE-LEARNING} {ALGORITHM},'' \emph{The Astrophysical Journal}, vol. 798,
  no.~2, p. 135, Jan. 2015.

\bibitem{Bobra2014}
M.~G. Bobra, X.~Sun, J.~T. Hoeksema, M.~Turmon, Y.~Liu, K.~Hayashi, G.~Barnes,
  and K.~D. Leka, ``The helioseismic and magnetic imager ({HMI}) vector
  magnetic field pipeline: {SHARPs} {\textendash} space-weather {HMI} active
  region patches,'' \emph{Solar Physics}, vol. 289, no.~9, pp. 3549--3578, Apr.
  2014.

\bibitem{Bhattacharjee2020}
S.~Bhattacharjee, R.~Alshehhi, D.~B. Dhuri, and S.~M. Hanasoge, ``Supervised
  convolutional neural networks for classification of flaring and nonflaring
  active regions using line-of-sight magnetograms,'' \emph{The Astrophysical
  Journal}, vol. 898, no.~2, p.~98, Jul. 2020.

\bibitem{Yi2021}
K.~Yi, Y.-J. Moon, D.~Lim, E.~Park, and H.~Lee, ``Visual explanation of a deep
  learning solar flare forecast model and its relationship to physical
  parameters,'' \emph{The Astrophysical Journal}, vol. 910, no.~1, p.~8, Mar.
  2021.

\bibitem{gradcam}
R.~R. Selvaraju, M.~Cogswell, A.~Das, R.~Vedantam, D.~Parikh, and D.~Batra,
  ``Grad-{CAM}: Visual explanations from deep networks via gradient-based
  localization,'' in \emph{2017 {IEEE} International Conference on Computer
  Vision ({ICCV})}.\hskip 1em plus 0.5em minus 0.4em\relax {IEEE}, Oct. 2017.

\bibitem{gbackprop}
J.~T. Springenberg, A.~Dosovitskiy, T.~Brox, and M.~Riedmiller, ``Striving for
  simplicity: The all convolutional net,'' 2014.

\bibitem{Sun2022}
Z.~Sun, M.~G. Bobra, X.~Wang, Y.~Wang, H.~Sun, T.~Gombosi, Y.~Chen, and
  A.~Hero, ``Predicting solar flares using {CNN} and {LSTM} on two solar cycles
  of active region data,'' \emph{The Astrophysical Journal}, vol. 931, no.~2,
  p. 163, Jun. 2022.

\bibitem{Deeplift}
A.~Shrikumar, P.~Greenside, and A.~Kundaje, ``Learning important features
  through propagating activation differences,'' 2017.

\bibitem{IntGrad}
M.~Sundararajan, A.~Taly, and Q.~Yan, ``Axiomatic attribution for deep
  networks,'' 2017.

\bibitem{pandeyecml2023}
\BIBentryALTinterwordspacing
C.~Pandey, R.~A. Angryk, and B.~Aydin, ``Explaining full-disk deep learning
  model for solar flare prediction using attribution methods,'' 2023. [Online].
  Available: \url{https://arxiv.org/abs/2307.15878}
\BIBentrySTDinterwordspacing

\bibitem{Schou2011}
J.~Schou, P.~H. Scherrer, R.~I. Bush, R.~Wachter, S.~Couvidat, M.~C.
  Rabello-Soares, R.~S. Bogart, J.~T. Hoeksema, Y.~Liu, T.~L. Duvall, D.~J.
  Akin, B.~A. Allard, J.~W. Miles, R.~Rairden, R.~A. Shine, T.~D. Tarbell,
  A.~M. Title, C.~J. Wolfson, D.~F. Elmore, A.~A. Norton, and S.~Tomczyk,
  ``Design and ground calibration of the helioseismic and magnetic imager
  ({HMI}) instrument on the solar dynamics observatory ({SDO}),'' \emph{Solar
  Physics}, vol. 275, no. 1-2, pp. 229--259, Oct. 2011.

\bibitem{Pesnell2011}
W.~D. Pesnell, B.~J. Thompson, and P.~C. Chamberlin, ``The solar dynamics
  observatory ({SDO}),'' \emph{Solar Physics}, vol. 275, no. 1-2, pp. 3--15,
  Oct. 2011.

\bibitem{Muller2009}
D.~Muller, B.~Fleck, G.~Dimitoglou, B.~Caplins, D.~Amadigwe, J.~Ortiz,
  B.~Wamsler, A.~Alexanderian, V.~Hughitt, and J.~Ireland, ``{JHelioviewer}:
  Visualizing large sets of solar images using {JPEG} 2000,'' \emph{Computing
  in Science {\&} Engineering}, vol.~11, no.~5, pp. 38--47, Sep. 2009.

\bibitem{alex}
A.~Krizhevsky, ``One weird trick for parallelizing convolutional neural
  networks,'' 2014.

\bibitem{e23010018}
P.~Linardatos, V.~Papastefanopoulos, and S.~Kotsiantis, ``Explainable ai: A
  review of machine learning interpretability methods,'' \emph{Entropy},
  vol.~23, no.~1, 2021.

\bibitem{Occlusion}
M.~D. Zeiler and R.~Fergus, ``Visualizing and understanding convolutional
  networks,'' 2013.

\bibitem{Qiu2022}
L.~Qiu, Y.~Yang, C.~C. Cao, Y.~Zheng, H.~Ngai, J.~Hsiao, and L.~Chen,
  ``Generating perturbation-based explanations with robustness to
  out-of-distribution data,'' in \emph{Proceedings of the {ACM} Web Conference
  2022}.\hskip 1em plus 0.5em minus 0.4em\relax {ACM}, Apr. 2022.

\bibitem{Nielsen2022}
I.~E. Nielsen, D.~Dera, G.~Rasool, R.~P. Ramachandran, and N.~C. Bouaynaya,
  ``Robust explainability: A tutorial on gradient-based attribution methods for
  deep neural networks,'' \emph{{IEEE} Signal Processing Magazine}, vol.~39,
  no.~4, pp. 73--84, Jul. 2022.

\bibitem{Chattopadhay2018}
A.~Chattopadhay, A.~Sarkar, P.~Howlader, and V.~N. Balasubramanian,
  ``Grad-cam++: Generalized gradient-based visual explanations for deep
  convolutional networks,'' in \emph{2018 {IEEE} Winter Conference on
  Applications of Computer Vision ({WACV})}.\hskip 1em plus 0.5em minus
  0.4em\relax {IEEE}, Mar. 2018.

\bibitem{kokhlikyan2020captum}
N.~Kokhlikyan, V.~Miglani, M.~Martin, E.~Wang, B.~Alsallakh, J.~Reynolds,
  A.~Melnikov, N.~Kliushkina, C.~Araya, S.~Yan, and O.~Reblitz-Richardson,
  ``Captum: A unified and generic model interpretability library for pytorch,''
  2020.

\bibitem{sourcecode}
\BIBentryALTinterwordspacing
DMLab@GSU, ``{Open Source Repo: fdExplainGGCAM}.'' [Online]. Available:
  \url{{https://bitbucket.org/gsudmlab/fdexplainggcam/src/main/}}
\BIBentrySTDinterwordspacing

\bibitem{Tsurutani2006}
B.~Tsurutani, A.~Mannucci, B.~Iijima, F.~Guarnieri, W.~Gonzalez, D.~Judge,
  P.~Gangopadhyay, and J.~Pap, ``The extreme halloween 2003 solar flares (and
  bastille day, 2000 flare), {ICMEs}, and resultant extreme ionospheric
  effects: A review,'' \emph{Advances in Space Research}, vol.~37, no.~8, pp.
  1583--1588, Jan. 2006.

\bibitem{Toriumi2019}
S.~Toriumi and H.~Wang, ``Flare-productive active regions,'' \emph{Living
  Reviews in Solar Physics}, vol.~16, no.~1, May 2019.

\bibitem{Campi2019}
C.~Campi, F.~Benvenuto, A.~M. Massone, D.~S. Bloomfield, M.~K. Georgoulis, and
  M.~Piana, ``Feature ranking of active region source properties in solar flare
  forecasting and the uncompromised stochasticity of flare occurrence,''
  \emph{The ApJ}, vol. 883, no.~2, p. 150, Sep. 2019.

\bibitem{ji2021modular}
A.~Ji, A.~Arya, D.~Kempton, R.~Angryk, M.~K. Georgoulis, and B.~Aydin, ``A
  modular approach to building solar energetic particle event forecasting
  systems,'' in \emph{2021 IEEE Third International Conference on Cognitive
  Machine Intelligence (CogMI)}.\hskip 1em plus 0.5em minus 0.4em\relax IEEE,
  2021, pp. 106--115.

\end{thebibliography}
\end{document}